# Four Tails Problems for Dynamical Collapse Theories

**Kelvin J. McQueen**

*The primary quantum mechanical equation of motion entails that measurements typically do not have determinate outcomes, but result in superpositions of all possible outcomes. Dynamical collapse theories (e.g. GRW) supplement this equation with a stochastic Gaussian collapse function, intended to collapse the superposition of outcomes into one outcome. But the Gaussian collapses are imperfect in a way that leaves the superpositions intact. This is the tails problem. There are several ways of making this problem more precise. But many authors dismiss the problem without considering the more severe formulations. Here I distinguish four distinct tails problems. The first (bare tails problem) and second (structured tails problem) exist in the literature. I argue that while the first is a pseudo-problem, the second has not been adequately addressed. The third (multiverse tails problem) reformulates the second to account for recently discovered dynamical consequences of collapse. Finally the fourth (tails problem dilemma) shows that solving the third by replacing the Gaussian with a non-Gaussian collapse function introduces new conflict with relativity theory.*

*Keywords:* Quantum theory; State vector reduction; wave-function collapse; Tails Problem; GRW; Measurement Problem

**1. Dynamical Collapse Theories and the Problem of Outcomes**

Quantum mechanics suffers from the measurement problem. There are several ways of stating this problem. A useful formulation is the *problem of outcomes* (Maudlin 1995). Two propositions made plausible by quantum mechanics are prima facie incompatible with a third independently plausible proposition:

(A) The wave-function of a system is complete in the sense that the wave-function specifies (directly or indirectly) all of the physical properties of a system.

(B) The wave-function always evolves in accord with a linear dynamical equation (e.g. the Schrödinger equation).

(C) Measurements always (or at least usually) have single determinate outcomes i.e. at the end of the measurement the measuring device indicates a definite physical state.

Propositions (A)-(C) cannot all be true at once. To illustrate: let the wave-function of a (macroscopic) measuring device $d$ be described by $|ready>_d$ meaning that $d$ is ready to measure the physical state of some particle. Let $|0>$ and $|1>$ describe the distinct values of some two-valued property (e.g. spin); $d$ is designed to detect which of these two states a particle is in. If the wave-function for $d$ and particle $p$ is initially $|ready>_d|0>_p$ then switching $d$ on (i.e. performing the measurement) gives $|'0'>_d|0>_p$ where $|'0'>_d$ means that $d$ has detected that $p$ is in state $|0>$ (and displays this e.g. using a pointer). Similarly, if the initial wave-function is instead $|ready>_d|1>_p$ then turning the device on gives $|'1'>_d|1>_p$. Now let the initial wave-function of $p$ be the linear superposition $a|0>_p + b|1>_p$ such that $a$ and $b$ are nonzero and $|a|^2 + |b|^2 = 1$. Then the complete initial wave-function is:

$$|ready>_d \left(a|0>_p + b|1>_p\right) \qquad (1)$$

The linearity of the dynamics (B) guarantees that if $d$ is switched on then the microscopic superposition will be magnified up into the macroscopic object yielding an entangled superposition:

$$a|'0'>_d |0>_p + b|'1'>_d |1>_p \qquad (2)$$

Since $d$ only displays three possible states (ready, '0' and '1'), (2) does not represent a single definite measurement outcome. Thus, if propositions (A) and (B) are true then (given the kinds of physical states that actually obtain) proposition (C) cannot be true. But (C) is extremely plausible: it is apparently confirmed by observations of the post-measurement states of measuring devices. This is the problem of outcomes.

Solutions can be categorised in terms of which proposition they reject. Additional variables theories (e.g. Bohm (1952)) reject (A): the wave-function is incomplete and definite measurement outcomes are determined by additional variables. Orthodox textbook quantum mechanics denies (B): when measurement occurs the linear dynamics abruptly stops governing the measured system and the system's wave-function collapses into a definite (non-superposition) state. The probability that the wave-function collapses into one of its component states is given by the absolute value squared (*mod-square*) of that component's coefficient. The inadequacy of appealing to the vague notion of measurement is the traditional formulation of the measurement problem (Albert 1992: chapter 4).

The theories at issue - dynamical collapse theories - deny (B). Moreover, they solve the measurement problem by formulating the collapse process precisely, without reference to 'measurement' or cognates. I will illustrate dynamical collapse theories using the original GRW spontaneous collapse theory (Ghirardi et. al. 1986) and the modern matter-density theory (Ghirardi et. al. 1995). The latter denies (A) as well as (B), supplementing the wave-function with a matter-density distribution.[1]

But first it will be worth considering the Everett (or "many-worlds") interpretation,[2] which denies (C): the wave-function is complete and evolves only in accord with the linear equation. (1) and (2) are interpreted as a microscopic superposition causing a macroscopic system (*d*) to bifurcate into two distinct devices that report the two possible measurement outcomes. This does not contradict our experience because human observer *h* who is initially ready to observe *d*'s result (|ready>$_h$) will in accord with the linear dynamics branch too:

$$a|'0'>_h |'0'>_d |0>_p + b|'1'>_h |'1'>_d |1>_p \qquad (3)$$

The two separate terms in this superposition will (in realistic scenarios) undergo decoherence thereby evolving independently and will thus represent distinct macroscopic worlds within a single quantum mechanical multiverse.

Many-worlds theory is worth mentioning here because, as we shall see, the tails problem (when properly formulated) implies that the formalism of dynamical collapse theories describes a multiverse of some sort. This undermines collapse theories since their goal is to retain and explain (C) but (C) is inconsistent with many-worlds theory.

## 2. The GRW Theory

The idea behind the original GRW theory is simple: elementary particles have a tiny probability per unit time for spontaneously collapsing into a definite position. Measuring devices are composed of many entangled particles and so have an extremely high probability per unit time for collapsing into a definite position. In other words, although one isolated particle rarely spontaneously collapses, one non-isolated particle will certainly collapse if a particle it is entangled with collapses. GRW postulate that particles have a $10^{-16}$ probability per second for spontaneously collapsing.[3] Systems composed of $10^{23}$ entangled particles will then collapse around $10^7$ times per second. The probability that a given wave-function will collapse onto one of its components is given by the mod-square of that component's coefficient. This is how GRW recover the probabilistic predictions of textbook quantum mechanics. Note the two distinct roles played by probability: there is the probability per unit time for spontaneous collapse and there is the probability that the collapse will be centred on a given wave-function component.

---

[1] Ghirardi et. al. (1995: sec. 3) motivate the matter-density addition as a solution to the tails problem (see sec. 5.1 below). Meanwhile Allori et al. (2008: sec. 4.3) motivate the addition as a solution to the problem of the high (3N) dimensionality of the wave-function. Whether the dimensionality problem undermines (A) is controversial (Albert (2013)).
[2] Everett (1957), Saunders et al. (2010), Wallace (2012).
[3] In more sophisticated variants this holds for nucleons while electrons collapse more infrequently (Pearle and Squires: 1994).

It will be useful to distinguish the idealised GRW theory from the realistic GRW theory. Consider the following wave-function of a particle confined to the x-axis:

$$a_1|x_1> + a_2|x_2> + a_3|x_3> + \cdots + a_n|x_n> \qquad (4)$$

In the idealised theory this particle has a $10^{-16}$ probability per second for spontaneously collapsing into one of the components. The probability that the particle collapses to component $|x_i>$ is $|a_i|^2$. Now consider the transition from (1) to (2). In the idealised theory (2) is unstable and will immediately collapse into $|0'>_d|0>_p$ with probability $|a|^2$ or $|1'>_d|1>_p$ with probability $|b|^2$. This guarantees a definite measurement outcome.

Collapse in the idealised theory reduces all but one of the coefficients to zero, while the mod-square of the chosen coefficient goes to one. But this collapse function is unphysical.[4] This is due to position/momentum incompatibility. The more confined the position wave-function the more spread out the momentum wave-function. The more spread out the momentum wave-function the more equiprobable all possible states of momentum become. The relationship between energy and momentum then yields drastic post-collapse violations of energy conservation: ones that we know by experiment do not occur. So GRW formulated the collapse function as a Gaussian. The collapse then raises the mod-square of the chosen coefficient - the collapse centre - close to one while reducing the mod-square of all other coefficients close to zero but never actually to zero. GRW carefully chose the probability per unit time for spontaneous collapse and the width of the bell curve of the Gaussian ($10^{-5}$ meters), so that the energy conservation violations are consistent with known experiments.[5]

But in formulating the collapse function consistently with experiments, GRW may have undermined the theory's ability to explain definite measurement outcomes. We must now redefine what is meant by collapse. Collapsing into a wave-function component now means something like "shifting *most* of the mod-square value to that component". Reconsidering the transition from (1) to (2): (2) is unstable but there will be no transition to a state represented by one of the components as in the idealised theory. Rather, the post-collapse state is:

$$c|'0'>_d |0>_p + d|'1'>_d |1>_p \qquad (5)$$

---

[4] Even apart from the normalization problem for delta functions.
[5] For discussion see Feldman and Tumulka (2012).

Where $c$ and $d$ are nonzero, $|c|^2 + |d|^2 = 1$, and either $|c|^2 >> |d|^2$ or $|d|^2 >> |c|^2$. The probability that $|c|^2 >> |d|^2$ is $|a|^2$ while the probability that $|d|^2 >> |c|^2$ is $|b|^2$. To analyse the adequacy of this theory we must ask whether (5) ultimately makes sense as a description of a definite measurement outcome.

This theory is often denoted $GRW_0$ to distinguish it from $GRW_M$ (the matter-density formulation).[6] $GRW_M$ defines a three-dimensional matter-density distribution in terms of the wave-function using matter-density operators:

$$m(\boldsymbol{r}) = \sum_k m_k N^{(k)}(\boldsymbol{r}) \qquad (6)$$

$N^{(k)}(r)$ is a particle number operator that gives the number of particles of type k that exist at position r, and $m_k$ is the mass of a particle of type k. The matter-density function at r is then:

$$M(\boldsymbol{r}) = <\Psi|m(\boldsymbol{r})|\Psi> \qquad (7)$$

Here $|\Psi>$ is the quantum state vector of the universe. As Wallace (2008: 60) puts it, "the [matter] density is the sum of the mass-weighted 'probability' densities for finding each particle at r; for a one-particle wavefunction $\Psi(r)$, $M(r)$ is just $m$ times $|\Psi(r)|^2$". The wave-function $\Psi$ therefore determines the matter distribution $M(\boldsymbol{r})$. Originally the *matter*-density function was called the *mass*-density function. But as Allori et. al. (2014: 331-2) state: "the matter that we postulate in $GRW_M$ and whose density is given by the *m* function does not ipso facto have any such properties as mass or charge; it can only assume various levels of density".

I discuss the tails problems in the context of both $GRW_0$ and $GRW_M$ because these have become the most widely discussed collapse theories.[7] When I refer to GRW (without subscript) I am referring to $GRW_0$ and $GRW_M$. Furthermore, the meaning of "collapse centre" will depend on which GRW theory is under discussion. In $GRW_0$ "collapse centre" refers to the wave-function component that receives the amplitude increase, in $GRW_M$ "collapse centre" refers to the associated matter-density. We now turn to the tails problems.

---

[6] There is also $GRW_F$ (the flash ontology) on which matter is composed of discrete space-time points called "flashes". There is one flash for each collapse. The position of the flash is the position of the collapse centre and the time of the flash is the time of the collapse. See Bell (1987: chapter 22) and Allori et. al. (2008: sec. 3.2). I do not discuss $GRW_F$ since arguably, the collapse centre flashes are insufficient for the composition of macroscopic objects: Maudlin (2011: 257-8); Esfeld (forthcoming, sec. 3).

[7] Pearle's (1989) continuous spontaneous localisation (CSL) theory is superior to GRW because CSL respects the symmetry properties of the wave-function for identical constituents (Ghirardi et. al. 1995: 8). But since it makes no difference to the tails problem it is standard to discuss the issue in the simpler GRW theory (see e.g. Egg & Esfeld (forthcoming: sec. 3)).

## 4. Tails Problem I: The Bare Tails Problem

The traditional formulation of the tails problem is an inconsistent triad:

(i) The eigenstate-eigenvalue link (EEL)

(ii) The wave-function evolves in accord with the GRW dynamics.

(iii) Measurements have definite outcomes.[8]

EEL is the traditional interpretive principle of textbook quantum mechanics. It states that a system *S* has a measureable property *P* when and only when *S*'s quantum state is an eigenstate of *P*. In other words: *S* has some value of *P* if and only if *S* has some value of *P* with mod-square equal to one. But the GRW post-collapse state (5) is not an eigenstate of the property "displaying a single definite measurement outcome". That's because neither of the components have mod-square equal to one - one of the components is close to one while the other is close to zero. So we appear to have a problem.

GRW has three options:

Deny (i): retain the GRW Gaussian collapse function but revise EEL.

Deny (ii): retain EEL but revise the collapse function.

Deny (i) & (ii): revise both EEL and the collapse function.

Denying (i) has been the primary focus of the tails problem literature up until now, and will be discussed in a moment. Denying (ii) has not been explored, partly because there is no reason for retaining EEL (e.g. neither Bohm nor Everett retains EEL). I will argue that the tails problem (properly formulated) forces collapse theorists to deny (i) & (ii). Denying (iii) is not an option because explaining (iii) is the primary explanatory goal of collapse theories.

For simplicity let's focus on EEL applied to particle position: particle *p* is in region *R* if and only if the proportion of the total value of the mod-square associated with (points in) *R* is equal to one. How should we revise this principle? Since GRW post-collapse states come close to position eigenstates perhaps one only needs to relax EEL. This is the influential tactic of Albert and Loewer (1996) who postulate the so-called fuzzy-link:

**Fuzzy-link principle**: Particle *p* is in region *R* if and only if the proportion of the total mod-square value of *p*'s quantum state associated with points in *R* is greater than 1-*q*.

---
[8] See Albert & Loewer (1996) and Lewis (1995) for similar formulations.

Assuming this principle, it doesn't matter that (5) is not an eigenstate of the property "displaying a single definite measurement outcome". What matters is that there is some structure postulated by GRW, which is a suitable supervenience base for such a measurement outcome, and which is assigned mod-square value greater than 1-$q$. So problem solved: definite measurement outcomes are guaranteed by GRW's post-collapse states.

$GRW_M$ also appeals to the fuzzy-link. It just needs to be made clear that particle $p$ does not supervene on any aspect of $p$'s wave-function, but on the matter-density associated with the component of the particle's wave-function whose mod-square value is greater than 1-$q$.

According to Albert and Loewer the value of $q$ is defined to ensure that what we take to (not) be determinate is (not) determinate. However, a small continuum of $q$-values will yield this result equally well. The value of $q$ cannot be narrowed down any further by experiment. After all, $q$ is not a fundamental constant, so given an initial quantum state and the GRW dynamics, different hypotheses about the exact value of $q$ will not lead to different future quantum states.[9] For Albert and Loewer this is unsurprising given the metaphysical status of the fuzzy-link. They say that it "is a new proposal, an alternative proposal, about the relation between position-talk and quantum-talk; a new proposed supervenience rule" (1996: p87). Others have described it as "a proposal for how to use language" (Lewis: 2006).

Exactly how localised a wave-function needs to be for it to determine a localised particle is not well-defined due to $q$ being indeterminate. But for Albert and Loewer this kind of "vague supervenience" is nothing new: macroscopic objects vaguely supervene on microphysical particles (just as they always have) while microphysical particles (now) vaguely supervene on wave-functions (in $GRW_0$) and on matter-density distributions (in $GRW_M$). The problem looks solved. But there are more severe formulations the tails problem.

## 5. Tails Problem II: The Structured Tails Problem

An alternative statement of the tails problem was first formulated by Cordero (1999):

> "In my view, it cannot be a disturbing aspect of the GRW theory […] that it fails to preserve a (really mythical) value-eigenvalue link of standard quantum theory [...] There seem to be some relevant objections to the approach, however [...] the GRW approach *smuggles into the theory too many of the conceptual peculiarities of the "many decohering worlds" approach.*" (1999: 65)

---

[9] See Lewis (2003) for discussion.

This has been developed by Wallace (2008) who distinguishes the *bare* tails problem from what he calls the *structured* tails problem. Wallace formulates the latter as follows:

> "If the [fuzzy-link is] just a matter of descriptive convenience then what prevents us regarding observers as being just as present in the dead-cat term as in the live-cat term? After all [...] the dead-cat term is as rich in complex structure as the live-cat term [...] taken to its logical conclusion, this seems to suggest that the GRW theory [...] is just as much a 'many-worlds' theory as the Everett interpretation." (2008: 63)

Cordero (1999: 67) emphasises that $GRW_M$ does not escape this objection. Recall equations (6) and (7); the matter-density function effectively *defines* the tails *into* the ontology with the important caveat that low mod-square densities are *less dense*. The $GRW_M$ fuzzy-link states that a particle exists at a region if there is *enough* matter at that region. But whatever the non-zero density of the tails, they are still *structurally isomorphic* to the collapse centre.

I formulate the structured tails problem in terms of a conditional whose antecedent GRW must accept, *before* drawing the "Everett-in-disguise" moral from the consequent. This ensures that the formulation does not presuppose the Everett interpretation (or any of its interpretive principles). Recall that "collapse centre" refers to the randomly chosen wave-function component (in $GRW_0$) or the associated matter-density (in $GRW_M$). Similarly "tails" will refer to the tails of the wave-function (in $GRW_0$) or to the tails of the associated matter-density distribution (in $GRW_M$).

> **The Structured Tails Problem**: If the collapse centre structure determines a particle configuration, then so do the structures in the tails. This is because the tails and the collapse centre are structurally isomorphic (or at least relevantly structurally similar). Nothing about low mod-square value can suppress this isomorphic structure. The consequence is an Everettian many-worlds ontology.

To illustrate consider (5). But note that the way we wrote (5) begs the question against GRW. In particular GRW should deny that the terms $|'1'>_d$ and $|'0'>_d$ are a legitimate way to summarise the post-collapse wave-function. Writing the wave-function down in neutral terms gives:

$$c|[0-state]>_d |0>_p + \ d|[1-state]>_d |1>_p \qquad (8)$$

[0-state] is a complicated wave-function component structured so that *if* localised particles supervene on it (or its associated matter-density) then there are particles composing a measuring device displaying a 0-outcome.

[1-state] is an equally complicated wave-function component structured so that *if* localised particles supervene on it (or its associated matter-density) then there are particles composing a measuring device displaying a 1-outcome.

The central problem is that GRW tells us that *only* the collapse centre determines macroscopic objects without telling us *why*. It is not enough to say that high mod-square as opposed to low mod-square makes the difference. The theory must say why high mod-square values have this property. Otherwise there is a substantive explanatory hole in the theory reminiscent of the unexplained role of 'measurement' in textbook quantum mechanics. So given the structured tails problem it is unclear why GRW is an improvement on textbook quantum mechanics.

Albert and Loewer's idea of vague supervenience does not solve this problem. One cannot explain why our term "measuring device" picks out the high mod-square structure by asserting that "measuring device" is vague. The question is what our term "measuring device" *ought* to refer to given the fundamental description of reality given by GRW. And the argument here is that *if* the term refers to the high mod-square structure then it ought to refer to the low mod-square structure too for the sake of consistency: the two structures are isomorphic and nothing about mod-square value breaks the structural symmetry.

An analogy will help to press this point. The Everett interpretation suffers a similar explanatory problem: the problem of explaining why mod-square values are probabilities (Albert 2010). Thus, just as GRW has a problem explaining why high mod-square values are macro-existence determiners, Everettians have a problem explaining why high mod-square values are high probabilities. But there is a difference: Everettians admit this problem in the sense that they do not simply stipulate that high mod-square values are high probabilities as a matter of linguistic fiat. On the contrary, an enormous amount of work spanning decision theory and probability theory has gone into providing a solution (Saunders et. al. (2010), Wallace (2012)). The analogous problem in GRW is not typically addressed in this way as it is often said that the connection between high mod-square value and macro-existence can just be stipulated as a part of the GRW theory (e.g. via the fuzzy-link principle).

Not all GRW-advocates settle for the *ad hoc* stipulation. In fact there are a bewildering variety of responses without any apparent consensus on which is correct. Let us then consider some paradigmatic responses and consider whether they have responded specifically to the *structured* tails problem.

### 5.1 Clifton and Monton's response

Since mod-square values are probabilities, perhaps the solution is to analyse the concept of existence in terms of probability. Such a view is suggested by Clifton and Monton:

> "If one is willing to entertain the thought that events in a quantum world can happen without being mandated or made overwhelmingly likely by the wavefunction, then it is no longer clear why one should need to solve the measurement problem by collapsing wavefunctions! [...] ...one supposes there to be *a plausible intuitive connection between an event's having a high probability according to a theory, and the event actually occurring*".
>
> (Clifton and Monton 1999: 708 [My Italics])

Clifton and Monton motivate this view with the assertion that the connection between high probability and existence is "plausible", "intuitive", and "natural", and with the assertion that without this connection, collapse theories fail (Clifton and Monton 2000: 155).

Wallace (2008: 59) argues that Clifton and Monton confuse the role of probability in collapse theories. The mod-square of a superposition component is the objective probability for that component *to become a collapse centre*. It is not the objective probability for that component to have 'actually occurring' status. On a realist view of $GRW_0$ the entire wave-function exists (simpliciter) and on a realist view of $GRW_M$ the entire matter-density exists, including low-density tails.

Clifton and Monton could postulate that mod-square plays a *further* role beyond what GRW intended. This postulate would relate the existence of what composes macro-objects with mod-square values such that existence comes in degrees and there are borderline cases of whether or not something exists. In $GRW_M$ the mass-density would fade out of existence as its associated value drops below 1-q. And as the associated value goes above 1-q a higher level of existence (as well as density) is exemplified. Such a theory might conceivably be true. But much more work is needed to make sense of the idea of indeterminate existence, and to develop the theory more generally.[10]

**5.2 Ghirardi, Grassi and Benatti's response**

After introducing $GRW_M$, Ghirardi, Grassi and Benatti (1995) address the structured tails problem for $GRW_M$ with a further principle designed to define away the matter-density tails. They define the low-density matter in the tails as "inaccessible" (i.e. observers cannot directly measure it) and so "not objective". A problem with this response is described by Tumulka (2011: 6):

> "[The tails problem] is whether GRW theories give rise to unambiguous facts about the aliveness of Schrödinger's cat or the location of the marble [...] Such a worry cannot be answered by describing what an observer can or cannot measure. Instead, I think, the answer can only lie in what the ontology is like, not in what observers see of it."

---

[10] In an attempt to solve the Everettian probability problem, Vaidman (1998: 256) introduces the concept of the measure of existence of a world as a basis for introducing (an illusion of) probability.

Here is another problem: Ghirardi et al. cannot (without circularity) appeal to observers *until* they've solved the structured tails problem. After all, the observers in the tails can access the matter in the tails. So what's accessible to observers can only be defined by this theory after the structured tails problem has been solved. The fact that the observers in the collapse centre cannot access the matter-density in the tails does not appear to speak to the real tails problem.[11]

**5.3 Tumulka's and Albert's responses**

Tumulka (2011) defends $GRW_M$ against several objections that he thinks applies to $GRW_0$. He identifies the tails problem with the bare tails problem (see his sec. 3) but nonetheless offers a response to the structured tails problem in his discussion of Ghirardi et. al.'s idea of accessibility (sec. 5). Tumulka offers an argument by analogy. The analogy involves a marble in a box. The marble is associated with a wave-function with high mod-square value inside the box and a small patch outside the box. The question is whether the marble is in the box. Here is the argument:

> "[I]t is a fact for the matter-density ontology that the fraction $|c_2|^2 \ll 1$ of the marble's matter lies outside the box - a fact that does not contradict the claim that the marble is inside the box, as can be illustrated by noting that anyway, for thermodynamics reasons, the marble creates a vapor out of some of its atoms (with low partial pressure), an effect typically outweighing $|c_2|^2$. The state of the primitive ontology in which the overwhelming majority of matter is inside the box justifies saying that the marble is inside the box. Thus, the primitive ontology does provide a picture of reality that conforms with our everyday intuition."

The problem with this argument is that it only asks us to consider our everyday intuition regarding an extremely simple system. This is misleading. Our formulation of the structured tails problem requires that we consult our intuitions regarding more complex environments. For example, let

$$|ready>_d \left(a|in>_p + b|out>_p\right) \tag{9}$$

be the expression for *d* being ready to measure whether *p* is inside or outside the box such that $|a|^2 \gg |b|^2$. According to the GRW dynamics the subsequent state is probably:

$$a|'in'>_d |in>_p + b|'out'>_d |out>_p \tag{10}$$

We need (10) to bring out the concern about the low mod-square component. For $GRW_M$ the component $|'out'>_d$ means that there is a configuration of low-density matter that at least resembles a device indicating 'out' in its shape, dynamics, and function. Can we intuit this matter configuration as nothing more than small unimportant "bits" that are hanging off of the device that indicates 'in'? Imagine that the measuring device indicates an outcome using a pointer. How can the pointer pointing to 'in', have an array of parts (analogous to the vapour) that form the shape of a pointer pointing to 'out'? This does not seem intelligible.

---
[11] For further criticism see Monton (2004).

Maudlin (2010: 135-6) makes a similar point: "the low-density apparatus seems to have the same credentials to be a full-fledged macroscopic object as the high-density apparatus since the density per se does not affect the structural or functional properties of the object." Albert (2015: 150-4) has recently responded that this is "vividly and radically mistaken", that more careful analysis shows that low-density matter behaves quite differently, and that "once this is taken on board, there immediately ceases to be any such thing as a 'problem' about the tails".

Albert considers a billiard ball (ball 1) in a superposition of travelling to point P and travelling to point Q *from the left*, and another (ball 2) that is in a superposition of travelling to point P and travelling to point Q *from the right*. Prior to collision we have:

$$(a|\rightarrow P>_1 + b|\rightarrow Q>_1) \times (a|P \leftarrow>_2 + b|Q \leftarrow>_2) \qquad (11)$$

Where $|a|^2 >> |b|^2$. After collision we have:

$$a^2|\leftarrow P>_1 |P \rightarrow>_2 + b^2|\leftarrow Q>_1 |Q \rightarrow>_2 + \\ ab|Q \rightarrow>_1 |\leftarrow P>_2 + ab|P \rightarrow>_1 |\leftarrow Q>_2 \qquad (12)$$

Where $a^2|\leftarrow P>_1 |P \rightarrow>_2$ is the high density component in which two billiard balls, after a direction-of-motion reversing collision, are travelling away from P. Presumably *if* this component determines a post-collision state, then so does the $b^2$ component (while the remaining two determine situations in which collision was avoided). But Albert makes a distinction between the *high-density sector* and the *low-density sector*. In the high-density sector we have a familiar type of collision...

> "But look at the low-density sector: what happens there is that two balls converge at Q and pass right through one another - and (in the meantime) two new balls appear, which then recede, in opposite directions, from P." (2015: 154)

The implication is that high-density matter and only high-density matter has the structural and functional credentials to count as genuine macroscopic stuff.

The problem with this response is that the distinction between high and low density sectors is not fine-grained enough. If there were no reason to distinguish between distinct low-density sectors then Albert's conclusion follows and the problem is solved. But consider how Albert characterises low-density objects: "What are they? Call them (I don't know) ghosts" (154). Presumably "ghosts" is an apt description since *distinguishable objects* seem to be moving right through each other. Such objects are moving right through the high-density sector (and vice versa) too. Decoherence entails that in realistic circumstances there must be distinguishable sectors within the low-density sector, which are (to varying degrees) dynamically isolated from each other (despite overlapping in space). It is precisely this dynamic isolation that enables us to distinguish different low-density sectors. And as soon as we make such distinctions we immediately rediscover our structured tails, and the tails problem returns in earnest.

**5.4 Monton's and Chalmers' responses**

Monton (2004) responds to the problem of low-density tails (in $GRW_M$) by postulating psychophysical principles:

> "a certain assumption about psychophysical parallelism needs to be made. But the assumption is a reasonable one. [...] There is no reason to suppose that mental states supervene just on particle location; instead we can suppose that mental states supervene on the distribution of mass [matter]. Since the masses of particles in a brain are concentrated in the appropriate regions of space, it is reasonable to assume that the appropriate mental states supervene on those mass concentrations." (2004: 418)

The idea is that mental states supervene, not only on the *configurations* of matter-density in space-time, but also on the *denseness* of those configurations. And so because the matter-density tail that appears to compose a measuring device (pointing to 'out') and an observer (recording that result) has low-density, there are no experiences as of a measuring device (pointing to 'out'). That's because mental states require a brain-like structure to have sufficiently high matter-density. But on the face of it this doesn't help: we are left with tails containing objects that (other than being "low in density") are structurally isomorphic to humans but are unconscious. This is not an acceptable theory.

To avoid what might be called "zombie tails", one might supplement Monton's suggestion with an analysis of physical concepts such that their correct application to physical properties is partly determined by what experiences those physical properties induce in us. This option is explored by Chalmers (2012):

> "In the case of macroscopic spatial properties, it is plausible that spatial properties can be picked out by spatial concepts as that manifold of properties that serve as the causal basis for spatial experience [...] To simplify, the property of being two meters away from one might be picked out as the spatial relation that normally brings about the experience of being two meters away from one. [...] One can then argue that on a collapse interpretation, the properties and relations that normally bring about the relevant sort of spatial experiences are precisely properties and relations requiring the wavefunction's amplitude to be largely concentrated in a certain area." (2012: 295-296)

Combining this *phenomenal analysis* of physical concepts with Monton's psychophysical metaphysics removes the structured tails. For if low-density matter is never a causal basis for experiences of physical properties, then physical concepts do not apply to low-density matter. If we consider the marble in the box, the matter outside the box does not cause any experiences, only the matter inside the box does, and so 'marble' must refer only to what is inside the box. (Isolated particles cannot cause experiences, but they still have the *type* of property that cause the relevant experiences (high-density), which is all the phenomenal analysis needs to licence application of physical concepts to isolated particles.)

One problem with this proposal is that it is simply defining the problem away. Whatever we think of the phenomenal analysis of some physical concepts, there is no reason why all physical concepts must be so analysed. There can be concepts of matter, space and time with entirely non-phenomenal application conditions, which apply on the basis of what physical properties are instantiated. In that case, we can use these non-phenomenally analysed concepts to describe the tails, and to formulate the structured tails problem.

Another problem with this response is that it hides the tails problem behind the mind-body problem. There is no independent reason for supposing that density level is relevant to conscious experience. So this proposal is only tenable because we just don't know what physical properties determine conscious experience (the mind-body problem). If we can exploit the possibilities left open by the mind-body problem then it is not clear why spontaneous collapses should be postulated in the first place: just go back to the textbook theory, treat consciousness (rather than measurement) as the cause of collapse, blame any vagueness or incompleteness in this theory on the mind-body problem, and let that be the end of it.

### 5.5 Lewis' response

Lewis (2007: sec. 6) focuses on the structured tails problem and takes issue with the conclusion drawn here, that GRW is Everett-in-disguise. Lewis argues that the structured tails problem presupposes the correctness of the Everettian analysis of macroscopic superpositions (Wallace 2010). So the GRW advocate can simply say that because the Everett interpretation is false (e.g. due to the probability problem) the GRW ontology should not be interpreted in the Everettian manner (cf. Albert 2015: 160).

When Wallace (2003, 2010) defends the claim that *worlds* are determined by decohering wave-function branches he appeals to Dennett's criterion: "A macro-object is a pattern, and the existence of a pattern as a real thing depends on the usefulness [...] of the theories which admit that pattern in their ontology" (2003: 93). Thus, since there is a *pattern* resembling a measuring device in both the 'in' and 'out' components, both components contain measuring devices. According to Lewis, GRW need only reject this criterion:

> "The GRW theory [...] requires rejecting Dennett's criterion, since macro-objects are essentially tied to a particular kind of microphysical structure, namely high-amplitude structure. [...] But since the many-worlds theory also violates apparently obvious principles, the violation of Dennett's criterion is not in itself a reason to exclude the GRW theory from the set of solutions to the measurement problem." (Lewis 2007: 800).

In response: the conditional formulation of the structured tails problem dispels the claim that the formulation relies on Dennett's criterion. The premises of the argument that underlie our formulation are so minimal that we don't even assume that the collapse centre determines macro-existence. It is worth considering this point in detail.

If X determines Y then Y supervenes on X. By definition, if Y supervenes on X then there can be *no* changes in Y without changes in X. But there *can* still be changes in X without changes in Y. To illustrate: if a table supervenes on a classical particle configuration, then for the table to cease to exist there must be changes in the particles, but the table remains if one component particle ceases to exist. How many component particles must cease to be for the table to cease to be? The matter is indeterminate. But if we set aside problems arising from indeterminate or borderline cases (which need not concern us here) what we can say is that if two classical particle configurations are *sufficiently* isomorphic then if either configuration determines a table then both determine tables. More generally, if two elements of the fundamental ontology exhibit sufficient structural isomorphism then, if either determines a macro-structure then both determine structurally isomorphic macro-structures. From this *supervenience principle* we can construct our argument:

(i) If the collapse centre and the tails exhibit sufficient structural isomorphism then, if the collapse centre (or the tails) determines a macro-structure then the collapse centre and the tails determine structurally isomorphic macro-structures. [Supervenience principle]
(ii) If it is not the case that mod-square values explain differences in the macro-structures then the collapse centre and the tails exhibit sufficient structural isomorphism. [Explanation as our guide to supervenience]
(iii) It is not the case that mod-square values explain differences in the macro-structures. [From sections 5.1 - 5.4]
(iv) Hence, if the collapse centre (or the tails) determines a macro-structure then the collapse centre and the tails determine structurally isomorphic macro-structures. [From (i), (ii), & (iii)]

Like (i), (ii) is a very weak assumption: it simply assumes that mod-square values should explain why the tails don't determine macro-structures; otherwise we have no reason to believe that they make a difference to what supervenes on the GRW ontology. This is neutral on whether such an explanation consists in showing that the tails do or do not contain patterns that resemble familiar macro-structures. The survey from 5.1 to 5.4 strongly suggests that mod-square values fail to explain the relevant differences.

To derive (iv) we do not assume the Everett interpretation, or Dennett's criterion, or for that matter any criterion for illuminating supervenience relations. Indeed, at this stage we cannot even conclude that GRW determines *any* macro-structure let alone a multiverse structure. For that we must borrow a principle from GRW:

(v) The collapse centre determines a macro-structure. [GRW principle]
(vi) Hence, the collapse centre and the tails determine structurally isomorphic macro-structures. [From (iv) & (v)]

So if we grant GRW premise (v) then from quite minimal assumptions we can conclude that the GRW ontology is indistinguishable from Everett's.

The authors discussed from 5.1 to 5.4 would reject (iii). This enables them to deny the consequent of (ii). But Lewis does not take this route. Lewis wants to reject Dennett's criterion, but Dennett's criterion could only be relevant to (v) and rejecting (v) would amount to rejecting GRW. Lewis presumably would also not reject (i) since he poses no objection to the idea of supervenience, and implicitly relies on it with his phrase "essentially tied to". That leaves premise (ii).

Lewis may hold that mod-square values determine differences in the relevant macro-structures without explaining those differences. But this leads to a further problem concerning Lewis' analogy with Everett. Lewis suggests that just as Everettians must give up the intuitive principle that probability requires uncertainty, GRW must give up intuitive principles. But these are disanalogous because there is a substantive literature attempting to make sense of the concept of probability in terms of decision theory so as to *explain why* quantum probability does not require uncertainty. Since we demand this from Everettians (Lewis 2010), we should also demand from GRW an equally rigorous explanation of how mod-square values break the structural symmetry between collapse centre and tails.

A final problem is that Lewis' defence may actually pull the rug from under GRW, so to speak. For if we reject Dennett's criterion then we may have to deny premise (v). If macro-objects are not determined by the collapse centres in virtue of those centres exhibiting macro-object-like patterns, then why think there are any macro-objects at all in GRW? What's needed is *a theory* of how such structure comes together with high mod-square to yield macro-existence. But based on the survey through sections 5.1-5.4 it is not clear that any such theory is forthcoming.

## 6. Tails Problem III: The Multiverse Tails Problem

There are surprising ethical issues that depend on a resolution to the structured tails problem. The GRW collapse function actually distorts the tails making the GRW ontology distinguishable from the Everett ontology. Wallace (2014) demonstrates that the effect of collapse on the tail peak of a particle is to displace it towards the collapse centre. This can be seen by considering an equally weighted position superposition of two separated Gaussians (localised around positions x and y respectively). The collapse essentially multiplies this function by another Gaussian centred on (say) x. The effect on the Gaussian centred on y is not only to reduce it (to a tail) but to bring it closer to the Gaussian centred on x (the collapse centre). If the tail determines an atom then there will probably be atomic excitation. After some calculations of the effects of this Wallace concludes that radiation in the tails will increase to unhealthy levels:

> "With overwhelmingly high probability, agents will either observe quantum statistics very close to the averages predicted by quantum mechanics, or in due course die of radiation sickness." (2014: 4)

If this result is correct it motivates a slight revision to the formulation of the structured tails problem:

> **The Multiverse Tails Problem**: If the collapse centre structure determines a particle configuration, then so do the structures in the tails. This is because the tails and the collapse centre are structurally isomorphic (or at least relevantly structurally similar). Nothing about low mod-square value can suppress this isomorphic structure. The consequence is *a possible many-worlds* ontology.

The only difference here is italicised: we have replaced "Everett ontology" with "a possible many-worlds ontology". The idea is that the resulting ontology resembles a merely conceivable many-worlds ontology in which most worlds become overwhelmed with radiation. If this ontology is correct then one must consider the ethical implications of performing measurements that would generate such tails.[12] Despite their ultimate fate there are still measurement outcomes in the tails and so there is still sufficient conflict with the proposition that GRW aims to explain (C). This is the superior formulation of the most severe tails problem for GRW's Gaussian collapse function.

---

[12] In Everett there are also branches containing lethal radiation doses, though not necessarily as a direct result of our choices to perform measurements. Assuming the decision-theoretic arguments work Everettians are justified in ignoring these branches as improbable. This might suggest a way out for GRW: embrace the decision-theoretic arguments. But this would make collapses superfluous.

# 7. Tails Problem IV: The Tails Dilemma

A very different, and seldom explored response, reformulates the collapse function to kill off the multiverse tails before they emerge. This would require a compact support collapse function that retains the peak of the Gaussian (with its $10^{-5}$m width) but avoids the tails. This function would transform (4) by eliminating all of its components except those within (say) $10^{-5}$m of the collapse centre.

According to Lewis (1995) and Ghirardi (2011) compact support functions would not violate energy conservation in empirically disconfirmed ways. But regarding the tails problem, both papers argue that compact support functions fail to solve the *bare* tails problem (since position superpositions are guaranteed infinitely fast given collapse). Consequently both deny the need for compact support functions. But the tails problem at issue should be the *multiverse* tails problem.

Compact support collapse functions solve the multiverse tails problem as follows. Begin with a system described by (1). The system will likely never make it to (2) and if it does only for a fraction of a second. A particle's wave-function will undergo spontaneous collapse, localising all of the amplitude for every particle that composes the pointer, to within the region in which the pointer gives a definite result. The wave-function will instantaneously spread out to infinity. But these "tails" have no interesting structure and as soon as they build structure through entanglement a spontaneous collapse will occur again. Multiverses never get the chance to be created and the multiverse tails problem cannot arise.

But all is not well. For failing to solve the bare tails problem is not the only reason that collapse theorists reject compact support functions. To be taken seriously, collapse theories should be reconciled with relativity. The traditional challenge is that the collapse process apparently requires a preferred notion of distant simultaneity. While this challenge has arguably been addressed,[13] compact support functions introduce new challenges. I finish by outlining two such challenges.

The first comes from Pearle's (1997) discussion of the relativisation of his continuous spontaneous localistation (CSL) model. Pearle describes CSL's evolution equation (1997: sec. 4) which involves a term for the field that the state vector evolves under, which is crucial for the possibility of Lorentz transformations. Pearle then describes how tails are also essential for these transformations:

> "I want to give one more reason for tails: I can't see how to make a relativistic theory without them. If you have a tail, no matter how small, and you know the field which the state vector evolved under, you can run the evolution equation backwards and recover the statevector at any earlier time. If on the other hand, the tail was completely cut off, you get a nonsensical irrelevant earlier statevector. [...] One can go to another reference frame, and in doing so the frame sweeps backwards in time. I cannot see how you could get sensible results in another Lorentz frame without having the tail to tell you how to do it." (Pearle 1997: sec. 5.3).

---

[13] See Myrvold (2002). Bedingham (2011) and Simpson (2011) offer relativistic dynamical collapse theories.

The first challenge is therefore to show how one could possibly move between reference frames when the tails are being continually severed. The second challenge stems from the work of Hegerfeldt (1974, 1998), who argues that compact support localisation violates Einstein causality:

"If a particle were initially strictly localized in some region on earth and if there were a nonzero probability to observe the particle a fraction of a second later on the moon, this could be used for the absolute synchronization of clocks [...] if all systems were spread out over all space to begin with, then no problems would arise." (1998: sec. 4)

From the positivity of the Hamiltonian alone one can prove that tails are acquired infinitely fast if they are somehow destroyed even for a moment. This holds in relativistic quantum mechanics too. In their response to Hegerfeldt, Barat and Kimball (2003) clarify the formal structure of Hegerfeldt's theorem:

"Hegerfeldt's theorem is an "if...then" statement, and since the "then" part (immediate infinite tails) is nonsensical physics, we conclude that the "if" part of the theorem (localized wave functions) should not be realizable for a sound quantum theory. Thus Hegerfeldt's theorem really means that a logically consistent single-particle quantum theory should not allow localization." (2003: 110)

Another way to reach essentially the same conclusion is via the Reeh-Schlieder theorem, which entails that no state with bounded energy is an eigenstate of any observable associated with a bounded spacetime region (Haag 1996: theorem 5.3.2). Our fourth formulation of the tails problem is therefore the dilemma:

**The Tails Dilemma**: The collapse function is either formulated using a Gaussian collapse function that retains structured tails or a non-Gaussian (compact support) collapse function that eliminates them. Gaussian collapse functions are undermined by the multiverse tails problem. Non-Gaussian collapse functions are potentially undermined by the additional relativistic problems they give rise to.

## 8. Conclusion

Dynamical collapse theories that postulate Gaussian collapse functions are inadequate: they fail to explain determinate measurement outcomes (proposition (C)). In particular, such theories postulate that determinate measurement outcomes are grounded in particular structures within the wave-function (or associated matter-density). But the problem is that the wave-function (or the matter-density) is replete with similar structures. GRW therefore require a symmetry breaker to distinguish the favoured structures from the unfavoured structures. But the only relevant differences between them are their mod-square values, and there is no account that explains how such values could be the symmetry breakers. Collapse theories can be salvaged with compact support collapse functions, but then the additional relativistic problems need to be addressed.

**Acknowledgments**

I would like to thank Rachael Briggs, David Chalmers, Hilary Greaves, Daniel Nolan, Gabriel Rabin, Craig Savage, Wolfgang Schwarz, Michael Simpson, David Wallace, Alastair Wilson, and two anonymous referees, for helpful feedback. This publication was made possible in part through the support of a grant from Templeton World Charity Foundation. The opinions expressed in this publication are those of the author.